\documentclass[conference,10pt]{IEEEtran}
\usepackage{epsfig,rotating,setspace,latexsym,amsmath,epsf,amssymb,amsfonts,bm,theorem,subfigure,epstopdf}
\usepackage{cite,authblk}
\usepackage{bbm}

\usepackage{algorithm}
\usepackage[noend]{algpseudocode}

\algrenewcommand\algorithmicforall{\textbf{foreach}}
\algrenewcommand\algorithmicindent{.8em}

\usepackage{color}
\usepackage{mathtools}

\newtheorem{theorem}{Theorem}
\newtheorem{lemma}{Lemma}

\newenvironment{Proof}[1]{\medskip\par\noindent{\bf Proof:\,}\,#1}{{\mbox{\,$\blacksquare$}\par}}
\IEEEoverridecommandlockouts
\allowdisplaybreaks

\usepackage{wrapfig}

\begin{document}

	\title{Age of Gossip in Networks with \\ Community Structure \thanks{This work was supported by NSF Grants CCF 17-13977 and ECCS 18-07348.}}
    \author{Baturalp Buyukates \qquad Melih Bastopcu  \qquad Sennur Ulukus\\
	\normalsize Department of Electrical and Computer Engineering\\
	\normalsize University of Maryland, College Park, MD 20742\\
	\normalsize  \emph{baturalp@umd.edu} \qquad \emph{bastopcu@umd.edu}  \qquad \emph{ulukus@umd.edu}}
	
\maketitle

\begin{abstract}
We consider a network consisting of a single source and $n$ receiver nodes that are grouped into $m$ equal size communities, i.e., clusters, where each cluster includes $k$ nodes and is served by a dedicated cluster head. The source node keeps versions of an observed process and updates each cluster through the associated cluster head. Nodes within each cluster are connected to each other according to a given network topology. Based on this topology, each node relays its current update to its neighboring nodes by \emph{local gossiping}. We use the \emph{version age} metric to quantify information timeliness at the receiver nodes. We consider disconnected, ring, and fully connected network topologies for each cluster. For each of these network topologies, we characterize the average version age at each node and find the version age scaling as a function of the network size $n$. Our results indicate that per node version age scalings of $O(\sqrt{n})$, $O(n^{\frac{1}{3}})$, and $O(\log n)$ are achievable in disconnected, ring, and fully connected cluster models, respectively. Finally, through numerical evaluations, we determine the version age-optimum $(m,k)$ pairs as a function of the source, cluster head, and node update rates.
\end{abstract}
 
\section{Introduction}

Introduced in \cite{Kaul12a} to quantify timeliness in real-time status updating systems, the age of information metric has received significant attention across information, communication, networking, and queueing theory fields \cite{SunSurvey, YatesSurvey}. The classical age metric increases linearly in time in the absence of any updates and drops to a smaller value when an update is received. Thus, even if the information at the source does not change, the classical age at the receiver continues to increase as time passes, because of the underlying assumption that updates get stale with time. This may not necessarily be the case in many applications, including content delivery services. To remedy this, several variants of the classical age metric have been proposed in the literature, in which the age stays the same until the information at the source changes even if no updates are received. Among these are binary freshness metric \cite{Cho03}, age of synchronization \cite{Zhong18c}, and age of incorrect information \cite{Maatouk20}.

Similar in spirit, recently, a new age metric named \emph{version age} has appeared in the literature \cite{Yates21, Eryilmaz21}. Considering each update at the source as a version change, the version age counts how many versions out-of-date the information at a particular receiver is, compared to the version at the source. Version age increases by one when the source obtains fresher information, i.e., newer version. A predecessor of version age has appeared in \cite{Bastopcu20c}, which considers timely tracking of Poisson counting processes, that entails minimizing the count difference, i.e., version difference, between the process and its estimate.

Reference \cite{Yates21} characterizes the version age in memoryless gossip networks composed of $n$ arbitrarily connected nodes. To deliver information to the receiver nodes, the source in \cite{Yates21} employs a Poisson updating scheme (exponential inter-update times), as has been done previously in the context of social networks \cite{Ioannidis09}, timely tracking \cite{Bastopcu20c, Bastopcu21b}, and timely cache updating \cite{Bastopcu2021, Bastopcu20e, Kaswan21}. In addition to the updates arriving directly from the source, in \cite{Yates21}, nodes relay their update versions to each other. Also referred to as \emph{gossiping}, this network activity improves the age scaling since each node can receive updates from each other as well as from the source. In particular, \cite{Yates21} shows that the version age scales as $O(\sqrt{n})$ in a bi-directional ring network and as $O(\log n)$ in a fully connected network, where $n$ is the number of nodes. Earlier works on age scaling have considered the classical age metric and achieved $O(1)$ scaling in multicast networks \cite{Zhong17a, Zhong18b, Buyukates18, Buyukates18b, Buyukates19} using a centralized transmission scheme administered by the source, and $O(\log n)$ scaling in distributed peer-to-peer communication networks \cite{Buyukates21b} using a hierarchical local cooperation scheme.

\begin{figure}[t]
	\centerline{\includegraphics[width=0.88\columnwidth]{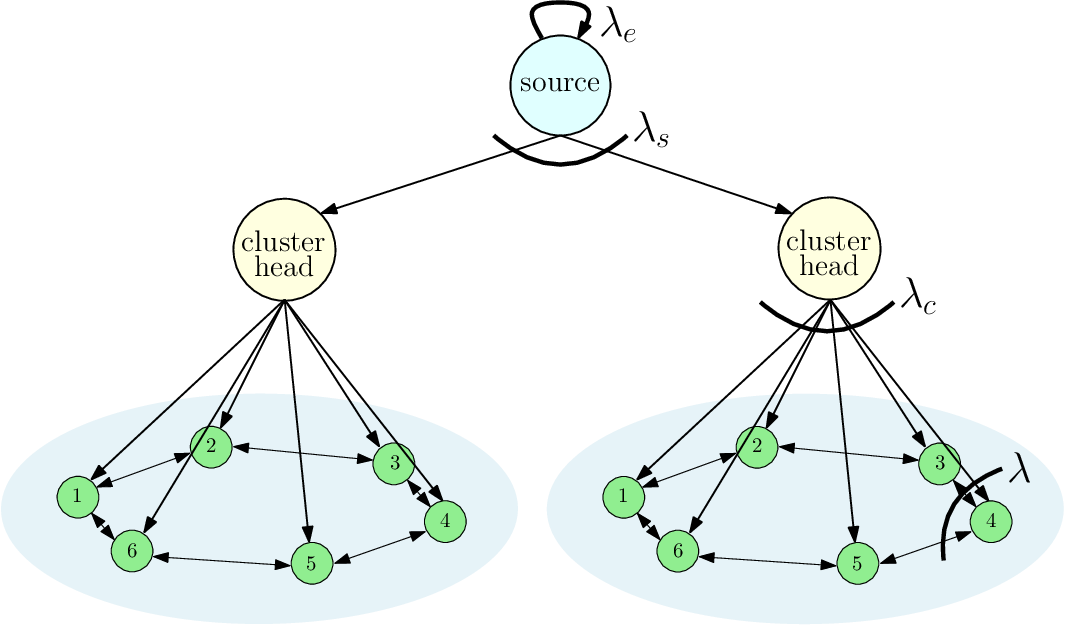}}
	\vspace{-0.15cm}
	\caption{Tiered network model where blue node at the center represents the source, yellow nodes represent the cluster heads, and green nodes represent the end users. Here, nodes in each cluster form a bi-directional ring network. Other possible network topologies within a cluster are shown in Fig.~\ref{Fig:netw_types}.}
	\vspace{-0.5cm}
	\label{Fig:disconnected}
\end{figure}

\begin{figure}[t]
	\centerline{\includegraphics[width=0.73\columnwidth]{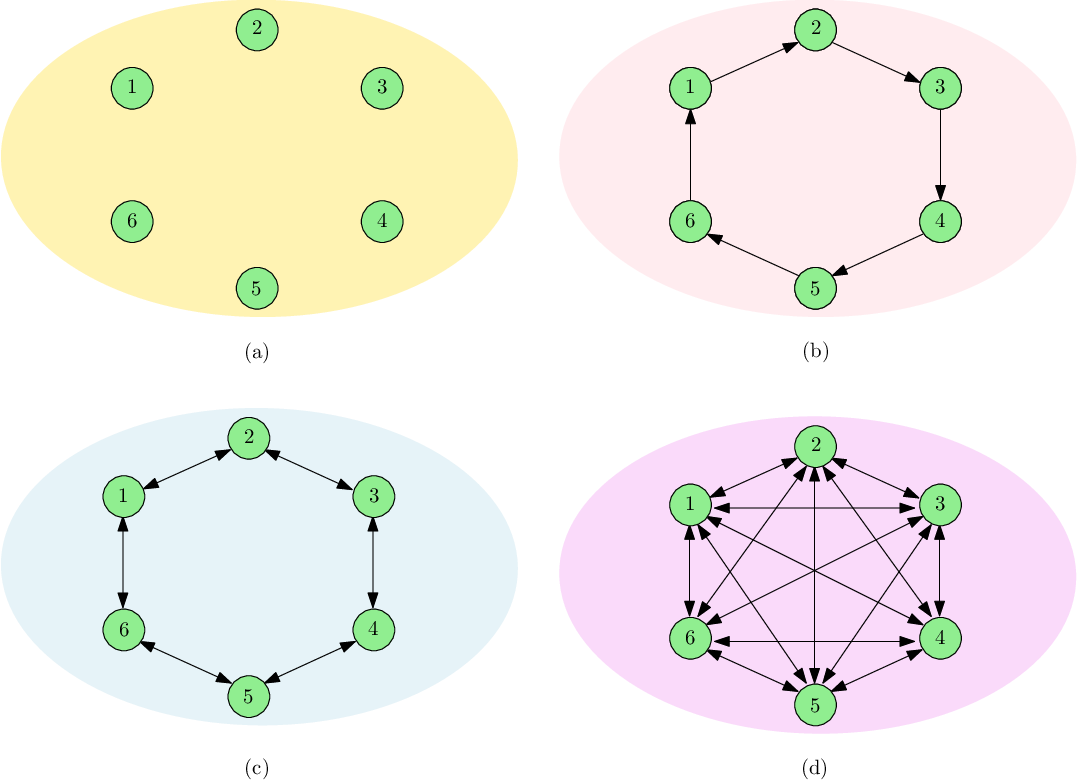}}
	\vspace{-0.2cm}
	\caption{Different network topologies that can be used within each cluster: (a) disconnected, (b) uni-directional ring, (c) bi-directional ring, and (d) fully connected. Fig.~\ref{Fig:disconnected} uses the one in (c).
	In this figure, cluster size is $k=6$.}
	\label{Fig:netw_types}
	\vspace{-0.5cm}
\end{figure}

Motivated by these, in this work, our aim is to investigate \emph{version age scaling} in more general gossip network models which exhibit a community structure; see Fig.~\ref{Fig:disconnected}. In our model, as in \cite{Yates21}, there is a single source that has an information which is updated following a Poisson process. Each such update at the source produces a newer version of the source information. The source node sends update packets regarding the source information to multiple communities. In our work, a community refers to a set of receiver nodes that are clustered together and can only interact with each other. Each cluster is served by a cluster head, which facilitates the communication with the source node. That is, cluster heads act as gateways between the source and the end-nodes within their clusters, akin to base stations in a cellular network. Unlike the model in \cite{Yates21}, in our model, the source cannot directly update individual nodes and updates arriving from the source go through the cluster heads. To model the various degrees of gossip within each cluster, we use disconnected, uni-directional ring, bi-directional ring, and fully connected network topologies; see Fig.~\ref{Fig:netw_types}. Based on the underlying connectivity within clusters, we characterize the version age experienced by each node. 

We observe that the additional hop constituted by the cluster heads between the source and the end-nodes presents us with opportunities to optimize the version age scaling by carefully tuning the number of clusters and the cluster size. Specifically, our results indicate that even if the nodes within each community forego gossiping, i.e., disconnected networks within each cluster, we can achieve $O(\sqrt{n})$ scaling as opposed to $O(n)$. In addition, we obtain the same $O(\log n)$ scaling in the case of fully connected communities using fewer connections within clusters than \cite{Yates21}, and further reduce the scaling result in ring networks to $O(n^{\frac{1}{3}})$ from $O(\sqrt{n})$ in \cite{Yates21}. Finally, through numerical evaluations, we determine the version-age optimum cluster sizes for varying update rates employed by the source, cluster heads, and the nodes within each cluster.

\section{System Model and the Age Metric}

We consider a system where a network of $n$ nodes is divided into $m$ clusters, each consisting of $k$ nodes such that $n = m k$ with $k, m \in \mathbb{Z}$; see Fig.~\ref{Fig:disconnected}. Each cluster is served by a distinct cluster head, which takes updates from the source and distributes them across that cluster. The source process is updated exogenously as a rate $\lambda_e$ Poisson process. The source has a total update injection rate of $\lambda_s$, which is uniformly distributed across cluster heads such that each cluster head is updated as a rate $\frac{\lambda_s}{m}$ Poisson process. From each cluster head to its corresponding cluster, the total update injection rate is $\lambda_{c}$ and this rate is uniformly allocated across the nodes in that cluster. That is, each node $i$ receives an update from its cluster head as a rate $\frac{\lambda_{c}}{k}$ Poisson process with $i \in \mathcal{N} = \{1, \ldots, n\}$.

Nodes in each cluster are connected to each other based on a connection graph. We consider varying levels of connectivity among nodes within each cluster. These are disconnected, uni-directional ring, bi-directional ring, and fully connected networks, which are shown in Fig.~\ref{Fig:netw_types} for a cluster of $k=6$ nodes. Updates received from the cluster head associated with each cluster are distributed across that cluster by utilizing the connections between the nodes. A node $i$ updates another node $j$ as a rate $\lambda_{ij}$ Poisson process. Each node in this system has a total update rate of $\lambda$, which is uniformly allocated to its neighboring nodes. That is, in the uni-directional ring, each node updates its neighbor node as a rate $\lambda$ Poisson process, whereas in bi-directional ring, each node has two neighboring nodes, each of which is updated as a rate $\frac{\lambda}{2}$ Poisson process. In the fully connected cluster, each node has $k-1$ neighbors each of which is updated as a rate $\frac{\lambda}{k-1}$ Poisson process. As a result of these local connections within a cluster, a node can receive different versions of the source update. 

To model the age at each node, we use the version age metric \cite{Yates21, Eryilmaz21}. We denote the version of the update at the source as $N_s(t)$, at cluster head $c$ as $N_{c}(t)$, with $c \in \mathcal{C} = \{1, \ldots, m\}$, and at node $i$ as $N_i(t)$, with $i \in \mathcal{N}$, at time $t$. The version age at node $i$ is given by $\Delta_i(t) = N_s(t)- N_i(t)$. Similarly, the version age at cluster head $c$ is  $\Delta_{c}(t) = N_{s}(t)- N_c(t)$. When node $i$ has the same version as the source, its version age becomes zero, i.e., $\Delta_i(t) = 0$. When the information at the source is updated, version ages at the cluster heads and the nodes increase by 1, e.g., $\Delta'_{c}(t) =\Delta_{c}(t)+1$. Each node $i$ can get updates either from its cluster head or the other nodes that it is connected to within its cluster. When node $i$ gets an update from its cluster head, its version age becomes $\Delta'_i(t) = \min\{\Delta_{c}(t),\Delta_i(t)\} = \Delta_{c}(t)$. Last equality here follows since nodes in a cluster receive the source updates through their cluster head so that they have either the same version or older versions of the information compared to their cluster head. When node $i$ receives an update from node $j$, its version age becomes $\Delta'_i(t) = \min\{\Delta_{i}(t),\Delta_j(t)\}$. We note that if node $j$ does not have a fresher version of the information, the version age at node $i$ is not updated. 

\section{Version Age with Community Structure}\label{sect:comm_age}

In this section, we characterize the limiting version age of each node $i$, denoted by ${\Delta}_i = \lim_{t \to \infty} \mathbb{E}[\Delta_i(t)]$, $i \in \{1, \ldots, n\}$ considering various network topologies for the clusters. Since the network model in each cluster is identical and within each cluster the network is symmetric for each of the network topologies, age processes $\Delta_i(t)$ of all users are statistically identical. Thus, in the ensuing analysis, we focus on a single cluster $c \in \mathcal{C}$ and find the average version age of a node from that cluster. For this, we follow the construction in \cite{Yates21} and express ${\Delta}_i$ in terms of ${\Delta}_S$, which denotes the average version age of subset $S$, where ${\Delta}_S(t) \triangleq \min_{j \in S} \Delta_j (t)$. 

We recall the following definitions from \cite{Yates21}: $\lambda_i(S)$ denotes the total update rate at which a node $i$ updates the nodes in set $S$. We have $\lambda_i (S) = \sum_{j \in S} \lambda_{ij}$ when $i \notin S$. Similarly, $\lambda_{c}(S)$ denotes the total update rate of the cluster head of a particular cluster into the set $S$. Finally, set of updating neighbors of a set $S$ is $N(S) = \{ i \in \{1,\ldots,n \} : \lambda_i (S) > 0 \}$.

With these definitions, next, in Theorem~\ref{thm_age_set} below we give the resulting version age in our clustered system model as a specialization of \cite[Thm.~1]{Yates21}.

\begin{theorem}\label{thm_age_set}
 When the total network of $n$ nodes is divided into $m$ clusters, each of which consisting of a single cluster head and $k$ nodes with $n = m k$, the average version age of subset $S$ that is composed of nodes within a cluster is given by
 \begin{align}
     {\Delta}_S = \frac{\lambda_e + \lambda_{c}(S){\Delta}_{c} + \sum_{i \in N(S)} \lambda_i(S) {\Delta}_{S\cup \{i\}} }{\lambda_{c}(S) + \sum_{i \in N(S)} \lambda_i(S)},\label{Yates_recursion}
 \end{align}
 with ${\Delta}_{c} = m\frac{\lambda_e}{\lambda_s}$. 
\end{theorem}

Proof of Theorem~\ref{thm_age_set} follows by applying \cite[Thm.~1]{Yates21} to our clustered network model and noting that updates arrive at the nodes through designated cluster heads.

\subsection{Version Age in Clustered Disconnected Networks}\label{Sect:disconn}

Nodes in a cluster are not connected to each other. Thus, the network is a two-hop multicast network, where the first hop is from the source to $m$ cluster heads, and the second hop is from each cluster head to $k$ nodes; combine Fig.~\ref{Fig:disconnected} with Fig.~\ref{Fig:netw_types}(a). Multihop networks have been studied in \cite{Zhong17a, Zhong18b, Buyukates18, Buyukates18b, Buyukates19} considering the classical age metric, where the source keeps sending update packets until they are received by a certain number of nodes at each hop. We do not consider such centralized management of updates, but let the source update the cluster heads as Poisson processes, and let cluster heads forward these packets to the nodes within their clusters as further Poisson processes.

Let $S_1$ denote an arbitrary $1$-node subset of a cluster. Subset $S_1$ is only connected to the cluster head, i.e., $N(S_1) = \emptyset$. Using the recursion given in (\ref{Yates_recursion}), we find
\begin{align}
    {\Delta}_{S_1} = {\Delta}_{c} + k\frac{\lambda_e}{\lambda_{c}} = m\frac{\lambda_e}{\lambda_s} + k\frac{\lambda_e}{\lambda_{c}},\label{version_age_disconnect}
\end{align}
where ${\Delta}_{S_1}$ denotes the version age of a single node from the cluster. When the network consists of two-hops, version age is additive, in that the first term in (\ref{version_age_disconnect}) corresponds to the first hop and equals to the version age at the cluster head, whereas the second term in (\ref{version_age_disconnect}) corresponds to the version age at the second hop between the cluster head and a node.

\begin{theorem}\label{corr_disconn}
  In a clustered network of disconnected users, the version age of a single user scales as $O(\sqrt{n})$.
\end{theorem}

Theorem~\ref{corr_disconn} follows by selecting $k=\sqrt{n}$ with $m = \frac{n}{k}=\sqrt{n}$ in (\ref{version_age_disconnect}) for fixed $\lambda_e$, $\lambda_s$, $\lambda_c$, which do not depend on $n$. Theorem~\ref{corr_disconn} indicates that when nodes are grouped into $\sqrt{n}$ clusters, an age scaling of $O(\sqrt{n})$ is achievable even though users forego gossiping. With the absence of cluster heads, i.e., when the source is uniformly connected to each of the $n$ users, the version age scaling of each disconnected user would be $O(n)$. By utilizing clusters, we incur an additional hop, but significantly improve the scaling result from $O(n)$ to $O(\sqrt{n})$. 
 
\subsection{Version Age in Clustered Ring Networks} \label{Sect:ring}

Nodes in each cluster form a ring network. We consider two types of ring clusters: uni-directional ring as shown in Fig.~\ref{Fig:netw_types}(b) and bi-directional ring as shown in Fig.~\ref{Fig:netw_types}(c).

First, we consider the uni-directional ring and observe that an arbitrary subset of $j$ adjacent nodes $S_j$ has a single neighbor node that sends updates with rate $\lambda$ for $j \leq k-1$. Each such subset $S_j$ receives updates from the cluster head with a total rate of $j\frac{\lambda_{c}}{k}$. Next, we use the recursion in (\ref{Yates_recursion}) to write
\begin{align}\label{version_age_ring_with_base}
  {\Delta}_{S_j} = \frac{\lambda_e+j\frac{\lambda_{c}}{k}{\Delta}_{c}+\lambda {\Delta}_{S_{j+1}}}{j\frac{\lambda_{c}}{k}+\lambda},  
\end{align}
for $j \leq k-1$ where ${\Delta}_{c}$ is the version age at the cluster head. We note that when $j=k$ the network becomes a simple two-hop network similar to that of Section~\ref{Sect:disconn} and we find ${\Delta}_{S_k} =m\frac{\lambda_e}{\lambda_s}+\frac{\lambda_e}{\lambda_{c}}$.

Next, we consider the bi-directional ring and observe that an arbitrary subset $S_j$ that consists of any adjacent $j$ nodes has two neighbor nodes, each with an incoming update rate of $\frac{\lambda}{2}$ for $j < k-1$. When $j=k-1$, $S_j$ has a single neighboring node that sends updates with a total rate $2\frac{\lambda}{2}=\lambda$. For $j\leq k-1$, the cluster head sends updates to subset $S_j$ with a total rate of $j\frac{\lambda_{c}}{k}$. With all these, when we apply the recursion in (\ref{Yates_recursion}), we obtain exactly the same formula given in (\ref{version_age_ring_with_base}). 
\begin{lemma}\label{corr_ring}
 Both uni-directional and bi-directional ring cluster models yield the same version age for a single node when each node in a cluster has a total update rate of $\lambda$.
\end{lemma}

Lemma~\ref{corr_ring} follows from the fact that either type of ring cluster induces the same recursion for an arbitrary subset of any adjacent $j$ nodes within a cluster as long as the total update rate per node $\lambda$ is the same. Thus, in the remainder of this paper, we only consider the bi-directional ring cluster model.

Before focusing on age scaling in a clustered network with a ring topology in each cluster, we revisit the ring network in \cite{Yates21}, and provide a proof of the $1.25 \sqrt{n}$ age scaling result observed therein as a numerical result. We show that the approximate theoretical coefficient is $\sqrt{\frac{\pi}{2}}=1.2533$. 

\begin{lemma}\label{yates_ring_proof}
  For the ring network model considered in \cite{Yates21}, the version age of a user scales as $\Delta_{S_1} \approx \sqrt{\frac{\pi}{2}}\frac{\lambda_e}{\lambda}\sqrt{n}$.
\end{lemma}

\begin{Proof}
From recursive application of \cite[Eqn.~(17)]{Yates21}, we obtain 
\begin{align}\label{yates_age_scaling_exp} 
    \Delta_{S_1} = \frac{\lambda_e}{\lambda}  \left(\sum_{i=1}^{n-1} a^{(n)}_i+a^{(n)}_{n-1}\right),
\end{align}
where $a^{(n)}_i$ is given for $i = 1,\ldots, n-1$ as
\begin{align} \label{defn_a_i}
a^{(n)}_i = \prod_{j=1}^{i}  \frac{1}{1+\frac{j}{n}}.
\end{align}
We note that $a^{(n)}_i$ decays fast in $i$, and consider $i=o(n)$,
\begin{align}
\!\!\!\!\!-\log(a^{(n)}_i) = \sum_{j=1}^{i} \log \left(1+\frac{j}{n}\right) \approx  \sum_{j=1}^{i} \frac{j}{n} = \frac{i (i+1)}{2n} \approx \frac{i^2}{n} \label{yates_age_scaling_exp2} \!\!
\end{align}
where we used $\log(1+x)\approx x$ for small $x$, and ignored the $i$ term relative to $i^2$. Thus, for small $i$, we have $a^{(n)}_i \approx e^{-\frac{i^2}{2n}}$. For large $i$, $a^{(n)}_i$ converges quickly to zero due to multiplicative terms in $\prod_{j=1}^{i} \frac{1}{1+j/n}$, and this approximation still holds. Thus, we have $\sum_{i=1}^{n-1} a^{(n)}_i \approx \sum_{i=1}^{n-1} e^{-\frac{i^2}{2n}}$. For large $n$, by using Riemann sum approximation with steps $\frac{1}{\sqrt{n}}$, we obtain  
\begin{align}\label{integral}
    \frac{1}{\sqrt{n}}\sum_{i=1}^{n-1} a^{(n)}_i \approx \frac{1}{\sqrt{n}}\sum_{i=1}^{n-1} e^{-\frac{i^2}{2n}} = \int_{0}^{\infty} e^{-\frac{t^2}{2}} \,dt=\sqrt{\frac{\pi}{2}}. 
\end{align}
Thus, we get $\sum_{i=1}^{n-1} a^{(n)}_i \approx\sqrt{\frac{\pi}{2}}\sqrt{n} $. By inserting this in (\ref{yates_age_scaling_exp}), we obtain the age scaling of a user as $\Delta_{S_1} \approx   \sqrt{\frac{\pi}{2}}\frac{\lambda_e}{\lambda}\sqrt{n}$.
\end{Proof}

Next, we focus on age scaling in a clustered network with a ring topology in each cluster. 
From recursive application of (\ref{version_age_ring_with_base}) along with ${\Delta}_{S_k}$, we obtain
\begin{align}\label{scaling_exact}
    \!\!\!\!\Delta_{S_1} =&\frac{\lambda_e}{\lambda}\left(\sum_{i=1}^{k-1} b^{(k)}_i \right)+\Delta_c\left(1-b^{(k)}_{k-1}\right)
    +\Delta_{S_k} b^{(k)}_{k-1},
\end{align}
where similar to (\ref{defn_a_i}), $b^{(k)}_i$ is given for $i=1,\ldots, k-1$ as
\begin{align} \label{defn_b_i}
b^{(k)}_i = \prod_{j=1}^{i} \frac{1}{1+\frac{j}{k}\frac{\lambda_c}{\lambda}}. 
\end{align}
When $k$ is large, $b^{(k)}_{k-1}$ goes to zero, and $\Delta_{S_1}$ in (\ref{scaling_exact}) becomes 
\begin{align} \label{ring_approximated_age_final}
    \Delta_{S_1}\approx \frac{\lambda_e}{\lambda} \left(\sum_{i=1}^{k-1} b^{(k)}_i \right)+\Delta_c \!\approx   \sqrt{\frac{\pi}{2}}\frac{\lambda_e}{\sqrt{\lambda\lambda_c}}\sqrt{k}+m\frac{\lambda_e}{\lambda_s},
\end{align}
where the second approximation follows as in the proof of Lemma~\ref{yates_ring_proof}. Terms in (\ref{ring_approximated_age_final}) are $O(\sqrt{k})$ and $O(m)$, respectively. In \cite{Yates21}, there is a single cluster, i.e., $m=1$ and $k=n$, and thus, the version age scaling is $O(\sqrt{n})$. In our model, by carefully adjusting the number of clusters and the cluster sizes, we can improve this $O(\sqrt{n})$ scaling result to $O(n^{\frac{1}{3}})$.
 
\begin{theorem}\label{ring_scaling}
  In a clustered network with a ring topology in each cluster, the version age of a single user scales as $O(n^{\frac{1}{3}})$.
\end{theorem}

Theorem~\ref{ring_scaling} follows by selecting $m=n^{\frac{1}{3}}$ with $k=\frac{n}{m}=n^{\frac{2}{3}}$ in (\ref{ring_approximated_age_final}) for fixed $\lambda_e$, $\lambda_s$, $\lambda_c$, $\lambda$, which do not depend on $n$.

\subsection{Version Age in Clustered Fully Connected Networks}\label{Sect:fully}

Nodes in each cluster form a fully connected network where each node is connected to all the other nodes within its cluster with rate $\frac{\lambda}{k-1}$. We find the version age for a subset of $j$ nodes $S_j$ in a cluster. Each such subset $j$ has $k-j$ neighbor nodes in addition to the cluster head associated with their cluster. Using the recursion given in (\ref{Yates_recursion}), we find
\begin{align}\label{version_age_fully_with_base}
  {\Delta}_{S_j} = \frac{\lambda_e+{\frac{j\lambda_{c}}{k}{\Delta}_{c}}+\frac{j(k-j)\lambda}{k-1} {\Delta}_{S_{j+1}} }{\frac{j\lambda_{c}}{k}+\frac{j(k-j)\lambda}{k-1}},  
\end{align}
for $j \leq k-1$, where ${\Delta}_{c}$ is equal to $m\frac{\lambda_e}{\lambda_s}$. The average version age of the whole cluster is ${\Delta}_{S_k} = \Delta_c + \frac{\lambda_e}{\lambda_c} = m \frac{ \lambda_e}{\lambda_s}+\frac{\lambda_e}{\lambda_{c}}$. 

Next, we present bounds for ${\Delta}_{S_1}$.

\begin{lemma}\label{lemma_fully}  
When $\lambda_{c}=\lambda$, in a clustered network with fully connected topology in each cluster, the version age of a single node satisfies
 \begin{align}\label{version_age_fully_with_base_simp_6}
   \frac{(k-1)^2+k}{k^2} \Delta_c+ \frac{\lambda_e}{\lambda}\left(\frac{k-1}{k}\sum_{\ell=1}^{k-1}\frac{1}{\ell}+ \frac{1}{k} \right)\nonumber \\ \leq {\Delta}_{S_1} \leq \Delta_c + \frac{\lambda_e}{\lambda} \left(\sum_{\ell=1}^{k}\frac{1}{\ell} \right). 
 \end{align}
\end{lemma}

\begin{Proof}
We use steps similar to those in the proof of \cite[Thm.~2]{Yates21} and also consider the additional hop from the source to the cluster heads. For $\lambda_{c} = \lambda$, we take $j=k-\ell$ and (\ref{version_age_fully_with_base}) becomes
\begin{align}\label{version_age_fully_with_base_simp_1}
  {\Delta}_{S_{k-\ell}} = \frac{\frac{1}{k-\ell}\frac{\lambda_e}{\lambda} + \frac{1}{k} \Delta_c + \frac{\ell}{k-1}{\Delta}_{S_{k-\ell+1}} }{\frac{1}{k}+\frac{\ell}{k-1}},
\end{align}
for $\ell \leq k-1$ and ${\Delta}_{S_k} = \Delta_c +\frac{\lambda_e}{\lambda}$, where $\Delta_c$ is the age at the cluster head. Defining $\hat{\Delta}_{S_\ell} \triangleq {\Delta}_{S_{k-\ell+1}}$, we get
\begin{align}\label{version_age_fully_with_base_simp_2}
  \hat{\Delta}_{S_{\ell+1}} = \frac{ \frac{1}{k-\ell}\frac{\lambda_e}{\lambda} + \frac{1}{k}\Delta_c +  \frac{\ell}{k-1}\hat{\Delta}_{S_\ell} }{\frac{1}{k}+\frac{\ell}{k-1}}.  
\end{align}
Next, one can show that $\hat{\Delta}_{S_{\ell+1}}$ satisfies the following
\begin{align}\label{version_age_fully_with_base_simp_3}
    \hat{\Delta}_{S_{\ell+1}} \leq \frac{ \frac{1}{k-\ell}\frac{\lambda_e}{\lambda} + \frac{1}{k}\Delta_c +
    \frac{\ell}{k}\hat{\Delta}_{S_{\ell}} }{\frac{1}{k}+\frac{\ell}{k}}.
\end{align}
Defining $\tilde{\Delta}_{S_\ell} \triangleq \frac{\ell}{k}\hat{\Delta}_{S_{\ell}}$ and plugging it in (\ref{version_age_fully_with_base_simp_3}), we get
\begin{align}\label{version_age_fully_with_base_simp_4}
    \tilde{\Delta}_{S_{\ell+1}} = \frac{\ell+1}{k}\hat{\Delta}_{S_{\ell}} \leq \frac{1}{k-\ell}\frac{\lambda_e}{\lambda} + \frac{1}{k}\Delta_c + \tilde{\Delta}_{S_{\ell}}.
\end{align}
Noting that $\tilde{\Delta}_{S_1} = \frac{\hat{\Delta}_{S_1}}{k} = \frac{{\Delta}_{S_k}}{k} = \frac{1}{k}\left(\Delta_c +\frac{\lambda_e}{\lambda}\right)$, we write
\begin{align}\label{version_age_fully_with_base_simp_5}
    \tilde{\Delta}_{S_k} \leq \Delta_c + \frac{\lambda_e}{\lambda} \left(\sum_{\ell=1}^{k}\frac{1}{\ell} \right).
\end{align}
Since $\tilde{\Delta}_{S_k} = \hat{\Delta}_{S_k} = {\Delta}_{S_1}$,  (\ref{version_age_fully_with_base_simp_5}) presents an upper bound to the version age of a single node. For the lower bound, we follow similar steps starting from (\ref{version_age_fully_with_base_simp_2}). Detailed steps are omitted here due to space limitations.
\end{Proof}

From (\ref{version_age_fully_with_base_simp_6}), we see that for large $n$ with $\lambda_c = \lambda$, the version age of a single node $\Delta_{S_1}$ satisfies
\begin{align}\label{approximation}
\Delta_{S_1} \approx m\frac{\lambda_e}{\lambda_s} + \frac{\lambda_e}{\lambda}\log k.
\end{align}

\begin{theorem}\label{fully_conn_scaling}
  In a clustered network with a fully connected topology in each cluster, the version age of a single user scales as $O(\log{n})$.
\end{theorem}

Theorem~\ref{fully_conn_scaling} follows in multiple different ways. For instance, it follows by selecting $m=1$ and $k =\frac{n}{m}=n$. That is, we have a single fully connected network of $n$ users as in \cite{Yates21}. Theorem~\ref{fully_conn_scaling} also follows by selecting $m=\log n$ and $k=\frac{n}{m}=\frac{n}{\log n}$. That is, we have $\log(n)$ fully connected clusters with $\frac{n}{\log n}$ users in each cluster. Thus, version age obtained under a smaller cluster size with less connections is the same as that obtained when all nodes are connected to each other. In particular, in our model with $m=\log n$, each node has $O(\frac{n}{\log n})$ connections in comparison to $O(n)$ in \cite{Yates21}. 

Finally, we note that, a recurring theme in the analysis of clustered networks is the fact that the version age at an end-node $\Delta_{S_1}$ is \emph{almost} additive in the version age at the cluster head $\Delta_c$ as seen in (\ref{version_age_disconnect}), (\ref{scaling_exact}), and (\ref{version_age_fully_with_base_simp_6}). It is \emph{exactly} additive in the case of disconnected clusters in (\ref{version_age_disconnect}).   

\begin{figure}[t]
 	\begin{center}
 	\subfigure[]{%
 	\includegraphics[width=0.49\linewidth]{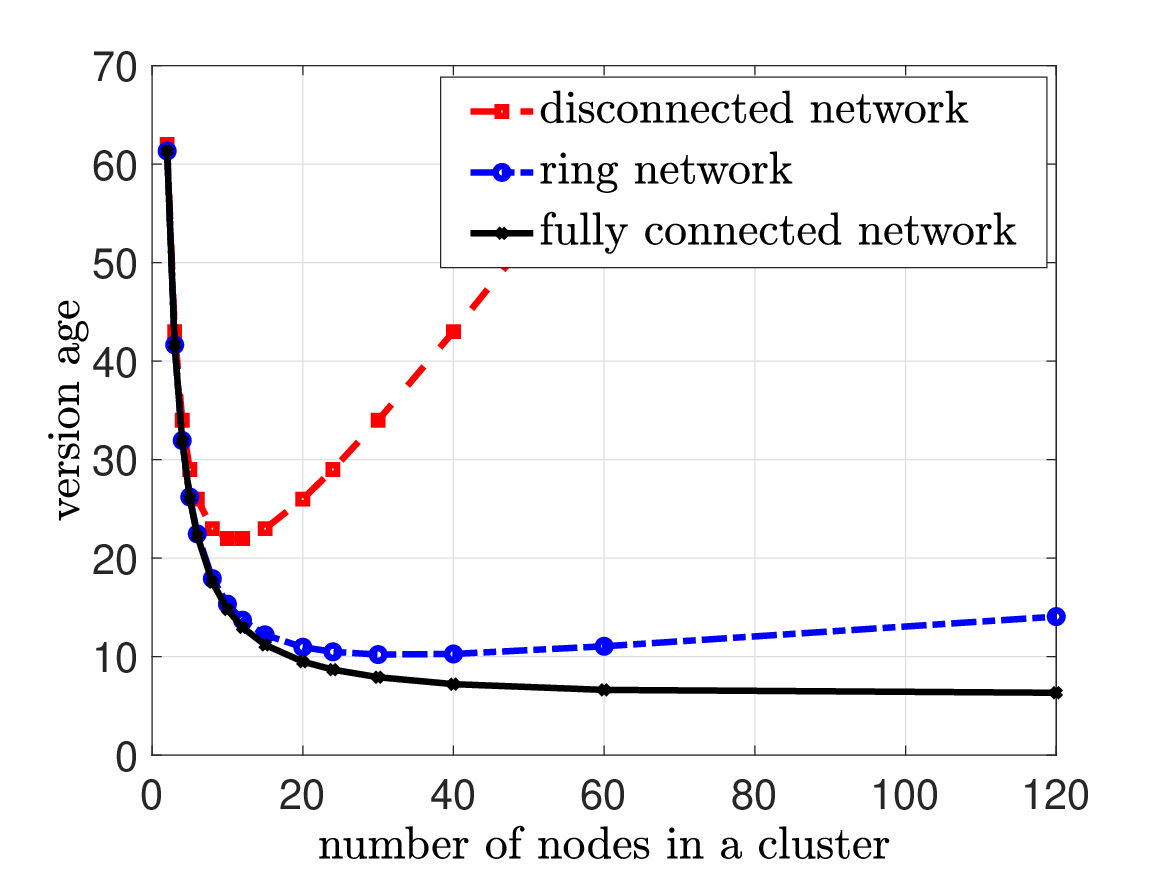}}
 	\subfigure[]{%
 	\includegraphics[width=0.49\linewidth]{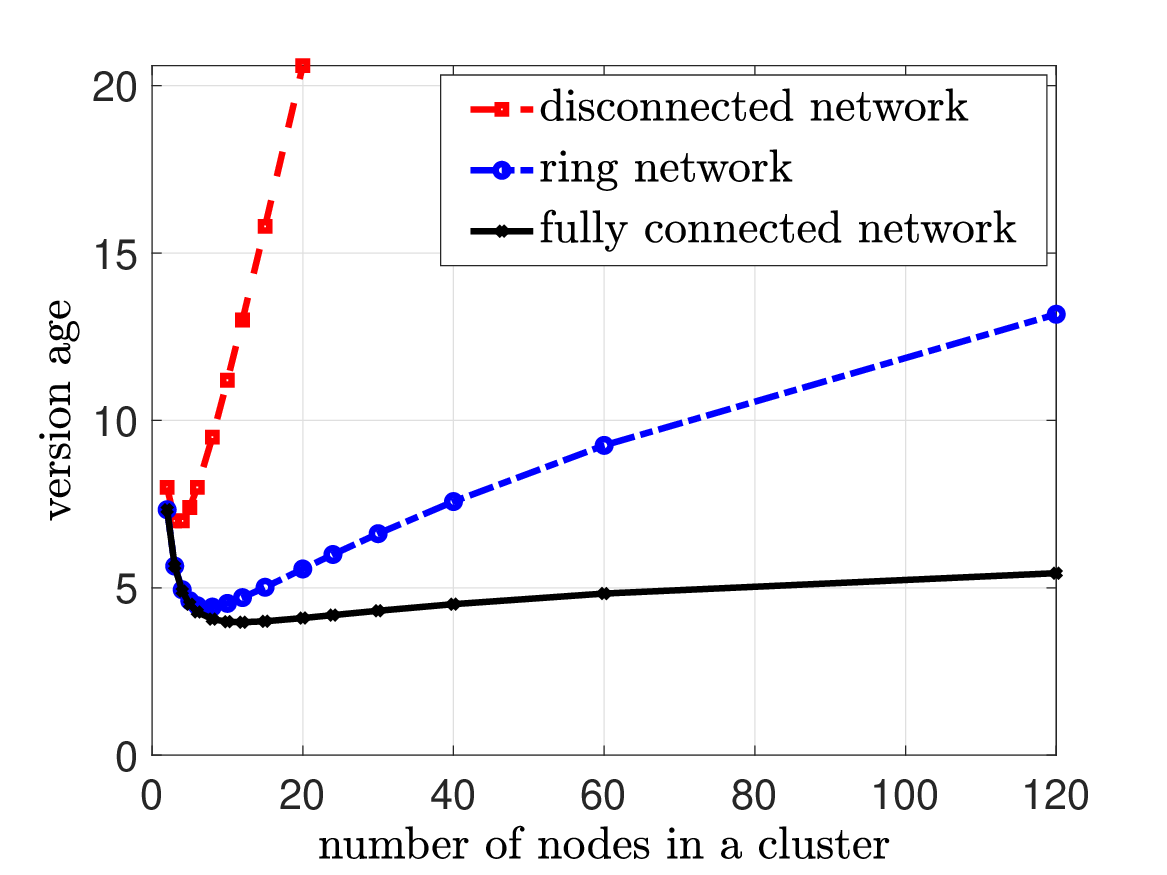}}\\ \vspace{-0.2cm}
 	\subfigure[]{%
 	\includegraphics[width=0.49\linewidth]{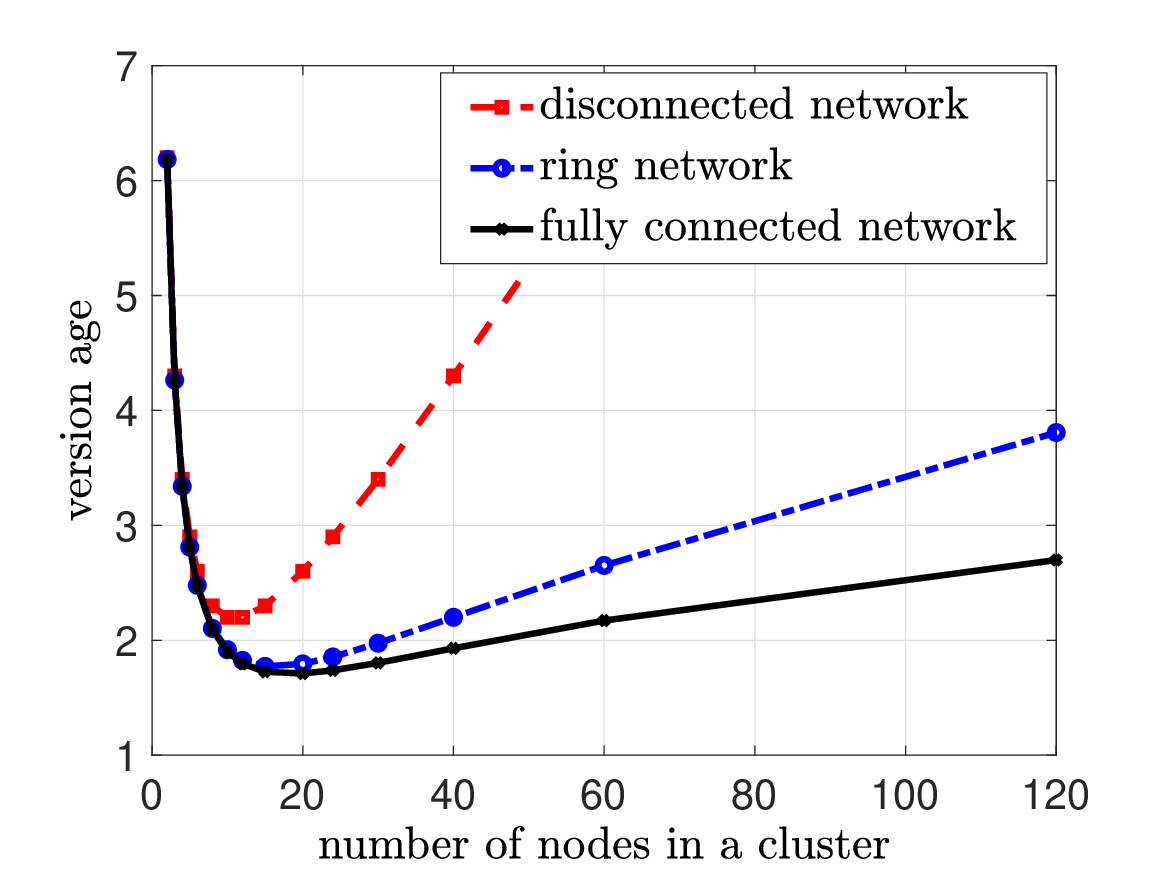}}
 	\subfigure[]{%
 	\includegraphics[width=0.49\linewidth]{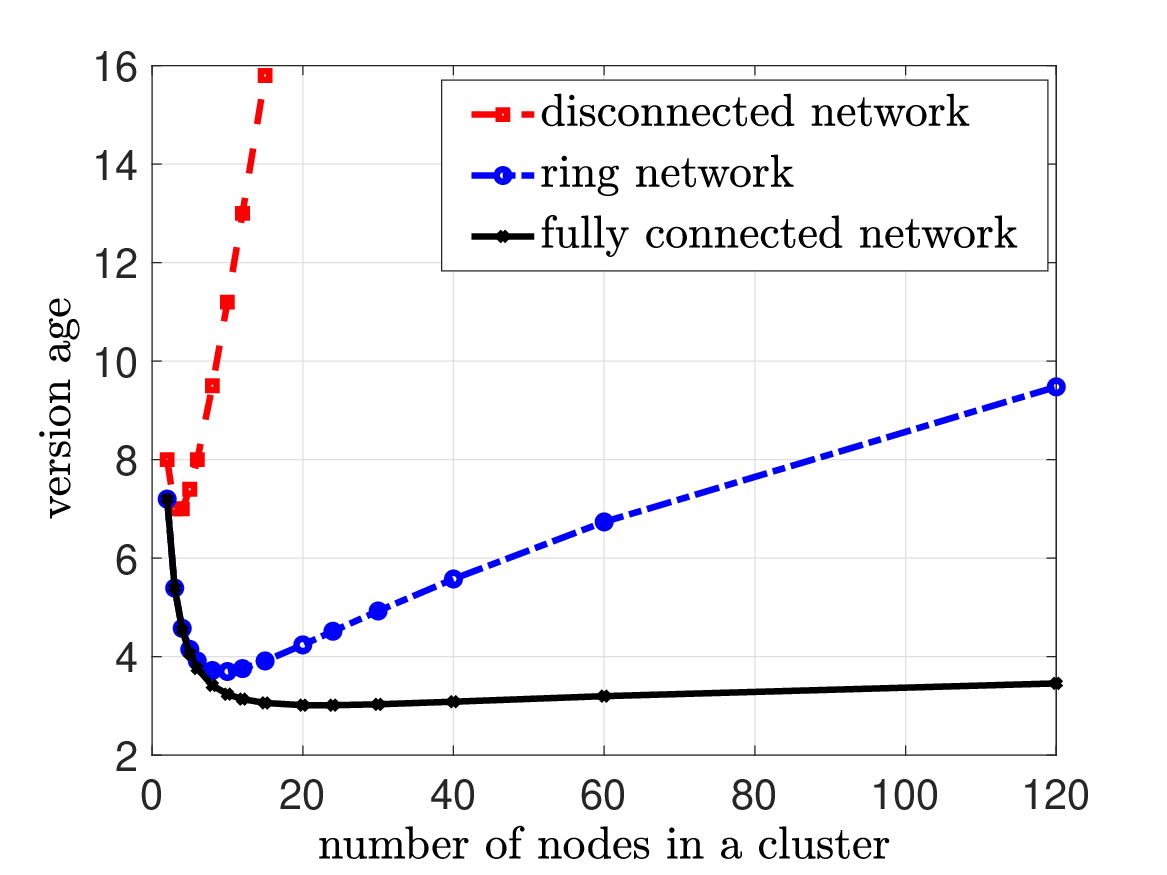}}
 	\end{center}
 	\vspace{-0.2cm}
 	\caption{Version age of a node with fully connected, ring, and disconnected cluster models with $n=120$, (a) $\lambda_e =1$, $\lambda_{s} =1$, $\lambda_{c} =1$, and $\lambda =1$, (b) $\lambda_e =1$, $\lambda_{s} =10$, $\lambda_{c} =1$, and $\lambda =1$, (c) $\lambda_e =1$, $\lambda_{s} =10$, $\lambda_{c} =10$, and $\lambda =1$, (d) $\lambda_e =1$, $\lambda_{s} =10$, $\lambda_{c} =1$, and $\lambda =2$. }
 	\label{Fig:sim_results_all_v2}
 	\vspace{-0.4cm}
\end{figure}

\section{Numerical Results}

We have seen in Section~\ref{sect:comm_age} that the version age depends on update rates $\lambda_e$, $\lambda_s$, $\lambda_c$, and $\lambda$. In this section, we explore the effects of these rates on the age via numerical results.

First, we take $\lambda_e =1$, $\lambda_{s} =1$, $\lambda_{c} =1$, $\lambda =1$, and $n=120$. We plot the version age of a node for the considered cluster models with respect to $k$. We see in Fig.~\ref{Fig:sim_results_all_v2}(a) that for the fully connected cluster model, the version age decreases with $k$ and thus, the version age-optimal cluster size is $k^*=120$, i.e., all $n$ nodes are grouped in a single cluster. In the ring cluster model, the version age is minimized when $k^*=30$. In the disconnected cluster model, the version age is minimized when we have $k^*=10$ or $k^*=12$. From these, we deduce that when the topology has less connectivity in a cluster, the optimal cluster size is smaller. Further, a topology with larger connectivity within a cluster achieves a lower version age.

Second, we consider the same setting as in Fig.~\ref{Fig:sim_results_all_v2}(a) but take $\lambda_s =10$ in Fig.~\ref{Fig:sim_results_all_v2}(b). Here, the version age decreases with increasing $k$ at first due to increasing number of connections within a cluster and the increase in the update rate between the source and each cluster head (as the number of clusters decreases with increasing $k$). However, as $k$ continues to increase, the decrease in the update rate from the cluster head to the nodes starts to dominate and the version age increases for all cluster models. In Fig.~\ref{Fig:sim_results_all_v2}(b), we see that the optimal cluster size is $k^*=12$ in fully connected clusters, $k^*=8$ in ring clusters, $k^*=3$ and $k^*=4$ in disconnected clusters.

Third, we increase the update rate of the cluster heads and take $\lambda_{c} =10$. We see in Fig.~\ref{Fig:sim_results_all_v2}(c) that the optimum value of $k$ increases compared to the second case when cluster heads have a larger update rate in all the cluster models. We find $k^*=20$ in fully connected clusters, $k^*=15$ in ring clusters, and $k^*=10$ or $k^*=12$ in disconnected clusters.  

Fourth, we study the effect of update rates among the nodes. For this, we take $\lambda_{c} =10$, $\lambda_e =1$, $\lambda_{s} =1$, $\lambda =2$. We see in Fig.~\ref{Fig:sim_results_all_v2}(d) that as the communication rate between the nodes increases, the optimal cluster size increases, and it is equal to $k^*=24$ in fully connected clusters, and $k^*=10$ in ring clusters. As there is no connection between nodes in the case of disconnected clusters, the optimum cluster size remains the same, i.e., $k^*=3$ or $k^*=4$, compared to Fig.~\ref{Fig:sim_results_all_v2}(b).

\bibliographystyle{unsrt}
\bibliography{IEEEabrv,lib_v6}
\end{document}